\documentclass[prl,showpacs,twocolumn]{revtex4}
\usepackage{amsmath}    
\usepackage{graphicx}   
\usepackage{verbatim}   
\usepackage{color}      
\usepackage{subfigure}  
\usepackage{hyperref}   
\usepackage{graphics}
\usepackage{epsfig}
\usepackage[english]{babel}
\usepackage{natbib}
\usepackage{fancybox}
\begin{document}

\title{Labyrinthine clustering in a spatial rock-paper-scissors ecosystem}

\author{Jeppe Juul, Kim Sneppen, and Joachim Mathiesen}

\affiliation{University of Copenhagen, Niels Bohr Institute, Blegdamsvej 17, DK-2100 Copenhagen, Denmark}

\date{\today}

\begin{abstract}
The spatial rock-paper-scissors ecosystem, where three species interact cyclically, is a model example of how spatial structure can maintain biodiversity. We here consider such a system for a broad range of interaction rates.
When one species grows very slowly, this species and its prey dominate the system by self-organizing into a labyrinthine configuration in which the third species propagates. The cluster size distributions of the two dominating species have heavy tails and the configuration is stabilized through a complex, spatial feedback loop. 
We introduce a new statistical measure that quantifies the amount of clustering in the spatial system by comparison with its mean field approximation. Hereby, we are able to quantitatively explain how the labyrinthine configuration slows down the dynamics and stabilizes the system.
\end{abstract}

\pacs{87.23.Kg, 87.23.Cc, 87.18.Hf}

\maketitle

\noindent \textbf{ \textit{ Introduction}}
--- Spatial migration of species is crucial for the viability of many ecological systems. As a striking example, crickets are known to locally deplete their nutritional resources to an extend where mass-migration is the only alternative to cannibalism \cite{simpson2006cannibal, sword2005insect}. Once the crickets have left an area, they can not return until the natural resources have been reestablished. Likewise, deadly viruses and bacteria depend on constantly infecting new hosts to survive \cite{morens2004challenge, sneppen2010minimal, juul2011locally}. 

\begin{figure}[tb]
    	\centering
            	\includegraphics[width=\columnwidth]{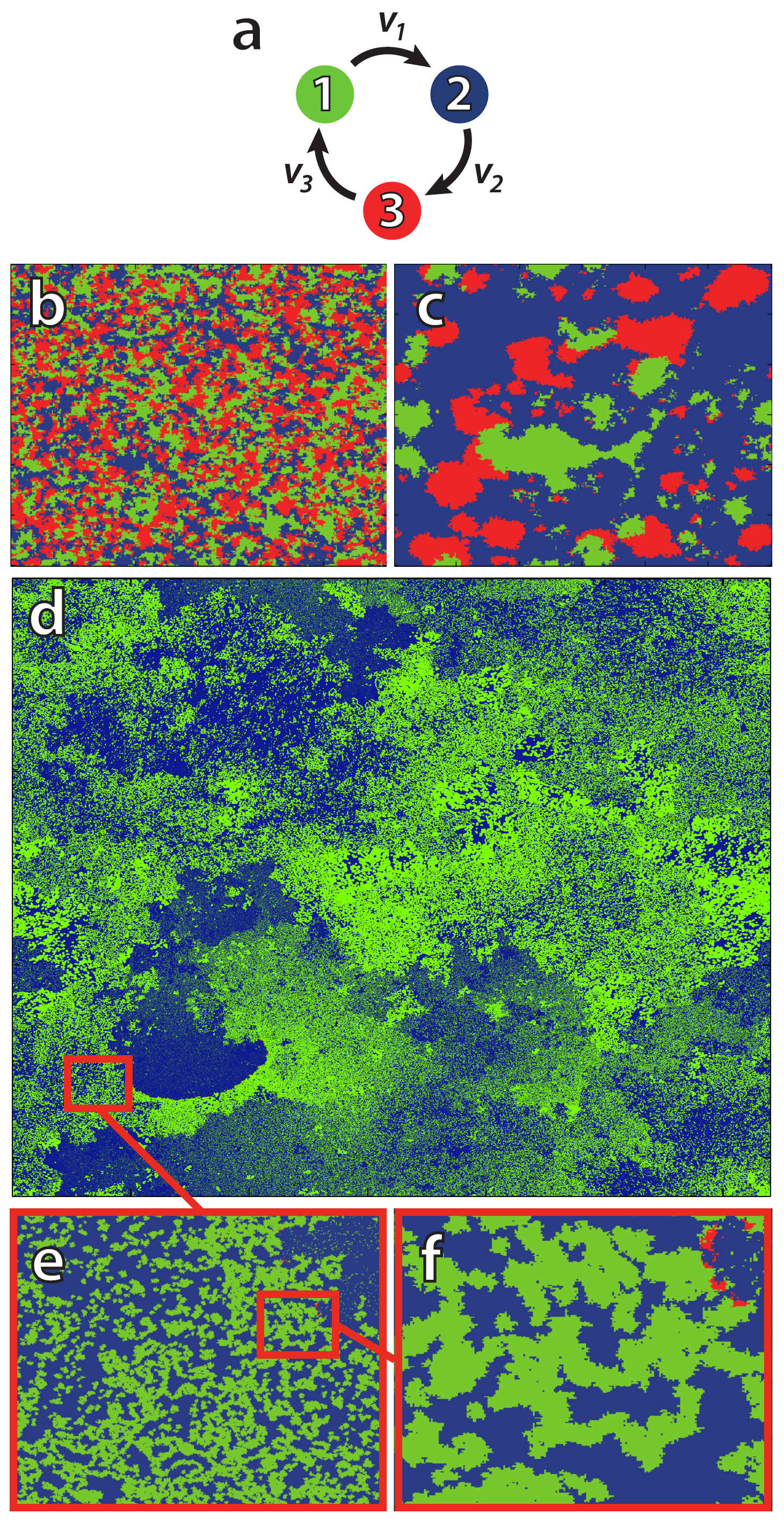}
    	\caption{(Color online) \textbf{Spatial self-organization in the rock-paper-scissors game.}
\textbf{a)} The three species interact cyclically. Species $i$ invades its prey at rate $v_i$.
\textbf{b-d)} Snapshots of the steady state spatial organization of the three species when 
\textbf{b)} All species grow at same rate and $L = 300$.   
\textbf{c)} Species 3 grows 5 times faster than 1 and 2 and $L = 300$.   
\textbf{d)} Species 1 grows 1250 times slower than 2 and 3 and $L = 12800$. 
\textbf{e-f)} zooms of the system in panel d.  
}
    	\label{fig:space}
\end{figure}

The rock-paper scissors game has emerged as a paradigm to describe the impact of spatial structure on biodiversity \cite{Frean, Kerr02, Reichenbach07, szabo2007evolutionary, frey2010evolutionary, avelino2012junctions}. In this system, three species interact cyclically such that species 1 can invade species 2, which can invade species 3, which, in turn, can invade species 1 (see Fig. \ref{fig:space}a). Such intransitive interaction pattern is very similar to the important genetic regulatory network \textit{the repressilator} \cite{elowitz2000synthetic, elowitz2002stochastic} and has been identified in many ecological system, among others in marine benthic systems \cite{jackson1975alleopathy, sebens1986spatial}, plant systems \cite{Cameron, Lankau, Taylor}, terrestial systems \cite{sinervo1996rock, Birkhead}, and microbial systems \cite{Durrett, Nahum, kirkup2004antibiotic, hibbing2010bacterial, trosvik2010web}. In such systems, all species constantly need to migrate spatially to survive.
Investigating three strands of \textit{E. coli} bacteria with cyclic interactions, it has been shown that biodiversity can not be preserved unless spacial structure is imposed by arranging the bacteria on a petri dish \cite{Kerr02, Kerr06, Reichenbach07}. These results have been reproduced in Monte Carlo simulations \cite{Frean, Johnson, Mathiesen, He10}, but even though many different analytical approaches have been applied, exactly how spatial structure stabilizes the system is still an open problem \cite{ Reichenbach06, Szabo04, Dobrinevski12, Juul12}.

\noindent \textbf{ \textit{  Model}}
--- We study the rock-paper-scissors game on a square lattice of $L \times L$ nodes and periodic boundary conditions. Each node is occupied by one of the three species 1, 2, or 3 growing at rates  $v_1$, $v_2$, and $v_3$, respectively. In each update a random node $i$ and a random of its neighbors $j$ are selected. If $i$ can invade $j$ according to the cyclic interacting pattern illustrated in Fig. \ref{fig:space}a, it will do so with a probability equal to $v_i$.

\noindent \textbf{ \textit{ Results}}
--- When the three species are initiated from a random configuration and with equal growth rates, they quickly organize into a steady state where all species are equally abundant and form small clusters (see Fig. \ref{fig:space}b). If the growth rate of species 3 is increased compared to species 1 and 2, species 2 becomes more abundant on the lattice and all three species form larger clusters (see Fig. \ref{fig:space}c). This paradoxical behavior, that the biomass of one species increases proportional to the growth rate of its prey, is characteristic for the rock-paper-scissors system  \cite{Frean, Johnson}.

Similarly, if the growth rate of species 1 is decreased, species 3 slowly becomes scarcer. Approaching the limit $v_1 \to 0$ a very large lattice is required in order for species 3 to be viable. In this limit a new, interesting spatial organization is observed. Species 3 propagates through the lattice in thin and broken wave fronts in constant flight from species 2. In the rest of the system the slowly growing species 1 and its prey, species 2, is tangled in a complex configuration with an enormous mutual perimeter. This spatial organization forms an ever-changing labyrinth of narrow pathways in which species 3 propagates (see Fig. \ref{fig:space}d-f). The more narrow and twisted the labyrinth becomes, the longer it will take for species 3 to return to a particular location, which gives species 1 more time to grow, forming broader pathways. This complex, spatial feedback loop stabilizes the configuration.

In order to describe this spatial self-organization mathematically, we study the probabilities $p_1$, $p_2$, and $p_3$ of a random node to be occupied by species 1, 2, or 3, respectively. Furthermore, we are interested in the perimeter $p_{ij}$ between species $i$ and $j$. That is, the probability of a random node and a random of its neighbors to be occupied by species $i$ and $j$, respectively. 

Given these perimeters the time evolution of species abundances is given by \cite{Reichenbach06}
 \begin{eqnarray}
        \dot{p}_1 = v_2 p_{12} - v_3 p_{31} , \label{pdot}
\end{eqnarray}
where the equations for $\dot{p}_2$ and $\dot{p}_3$ follow by cyclic permutation of the indices 1, 2, and 3. This symmetry also holds for all subsequent equations of this article.

\begin{figure}[tb]
    	\centering
            	\includegraphics[width=\columnwidth]{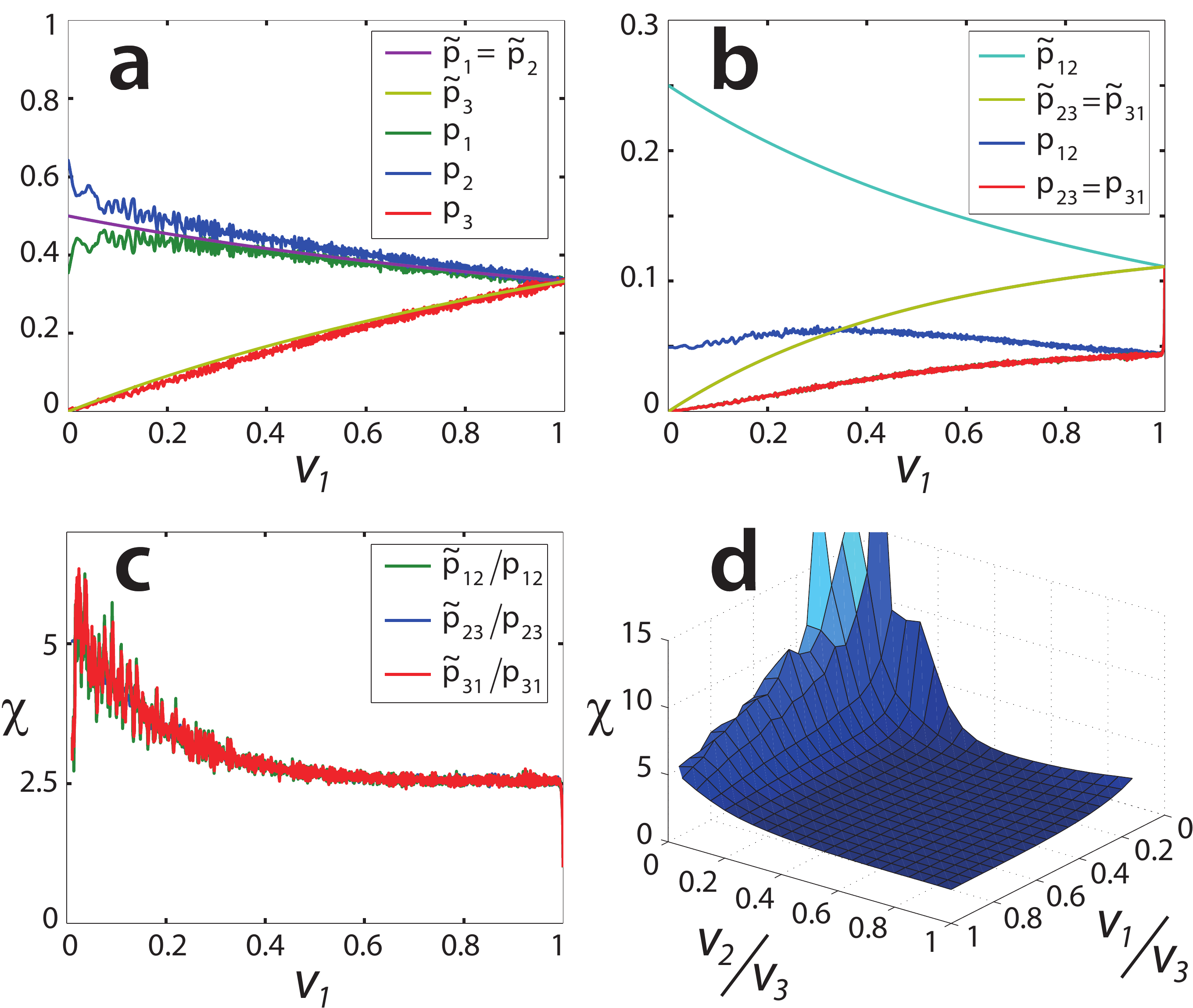}
    	\caption{(Color online) \textbf{Species abundancies and perimeters for small $v_1$.}
\textbf{a)} When the growth rate of species 1 is decreased, species 3 becomes less abundant. The mean field theory correctly predicts the abundancies.
\textbf{b)} The perimeters between the species are much lower than predicted by the mean field theory due to clustering.
\textbf{c)} The ratio between the predicted mean field perimeters and the observed perimeters are equal for all species. This ratio defines $\chi$.
\textbf{d)} When $v_3 \gg v_1, v_2$, the ratio $\chi$ diverges corresponding to the large clustering in Fig. \ref{fig:space}c. When one species grows much slower than the others $\chi$ approaches 5, which gives rise to the labyrinthine clustering in Fig. \ref{fig:space}d. 
}
    	\label{fig:pChi}
\end{figure}

To explain this behavior, one can adopt a mean field approximation, where all nodes are linked and spatial structure does not exist. Then, the perimeter between two species is simply given by the product of species abundances $\tilde{p}_{12} = \tilde{p}_1 \tilde{p}_2 $,
where tilde ($ \mathtt{\sim} $) denotes that the mean field approximation has been applied. 
If this is inserted into \eqref{pdot} and the time derivatives are set to zero, one obtains the steady state solution
\begin{eqnarray}
        \tilde{p}_1 = \frac{v_2}{v_1 + v_2 + v_3} \label{p1tilde} \\
        \tilde{p}_{12} = \frac{v_2 v_3}{(v_1 + v_2 + v_3)^2} \label{p12tilde} .
\end{eqnarray}
In Fig. \ref{fig:pChi}a-b the steady state abundances and perimeters are shown at constant $v_2=v_3=1$ and varying $v_1 \le 1$. It is seen that a slow growth rate of species 1 leads to a decline in the abundance of species 3 as expected. 
The mean field approximation correctly predicts how the abundances of the three species depend on the growth rates. However, the mean
 field approach can not capture the spatial organization of the species, and thus it predicts perimeters far longer than what is observed in simulations (see Fig. \ref{fig:pChi}b). The fact that the abundances are correctly predicted indicates that the mean field perimeters are proportional to the true, spatial perimeters. Indeed, if \eqref{pdot} is set to zero for both the spatial and mean field system one can derive the relations.
\begin{eqnarray}
        \frac{ \tilde{p}_{12} }{p_{12}}  = \frac{ \tilde{p}_{23} }{p_{23}}  = \frac{ \tilde{p}_{31} }{p_{31}}  \equiv \chi        \label{chi} .
\end{eqnarray}
Here, we have introduced the ratio $\chi$, defined by how much the perimeter between two species is longer in the steady state of the mean field approximation compared to the spatial system (see Fig. \ref{fig:pChi}c). This new statistical measure describes the spatial and dynamical organization of the rock-paper-scissors game for varying growth rates. The intuition behind $\chi$ is the following:

The average time before a node of species $2$ is invaded  by species 1 is given by $T_1 = \frac{p_1}{v_1 p_{12}}$ for the spatial system and $\tilde{T}_1 = \frac{\tilde{p}_1}{v_1 \tilde{p}_{12}}$ in mean field. Therefore, $\chi \approx \frac{T_1}{\tilde{T}_1}$ provides a measure for how much longer each species on average lives on a node before being invaded, compared to the mean field system, \textit{i.e.} how much the spatial organization slows down the dynamics.
Furthermore, when $\chi$ is large the perimeters of the spatial system is much smaller than in the mean field system, according to \eqref{chi}, so the species must have a high degree of clustering. Hence, $\chi$ gives a measure for the clustering of the spatial system. 
These two interpretations are, of course, tightly connected. If the average cluster diameters are doubled, each node will live for twice as long before being invaded, corresponding to increasing $\chi$ by a factor two.

How does $\chi$ depend on the growth rates of the three species? In Fig. \ref{fig:pChi}d this dependency is shown as a function of the relative growth rates $\frac{v_1}{v_3}$ and $\frac{v_2}{v_3}$, with $v_3$ chosen to be the fastest growing species. When all growth rates are equal we have $\chi \approx 2.5$, corresponding to the moderate amount of clustering observed in Fig. \ref{fig:space}b.  When species 3 grows much faster than the two other species, such that both growth ratios go to zero, $\chi$ becomes very large. This agrees well with the large amount of clustering observed in Fig. \ref{fig:space}c. 

When only $v_1 \to 0$ we see from Fig. \ref{fig:space}d and \ref{fig:pChi}d that $\chi$ approaches a finite value close to 5. In this limit, we expect from \eqref{p1tilde} that $p_3 \to 0$ while $p_1 \approx p_2 \to \frac 12$. In this case, it is limited how much species 3 can cluster. The observation of $\chi$ suggests that clustering only reduces the perimeter between species 2 and 3 by a factor 5 compared to the mean field system. This sets an upper bound for how much species 1 and 2 can cluster. 
The mean field approach predicts a perimeter $\tilde{p}_{12} = \frac 14$, so with $\chi \approx 5$ equation \eqref{chi} dictates the perimeter in the spatial system to be $p_{12} \approx \frac{1}{20}$. This agrees well with the $12800 \times 12800$ system in Fig. \ref{fig:space}d, where $v_1=0.0008$, $v_2 = v_3 = 1$, and $p_{12} \approx 0.05$, which is also evident from Fig. \ref{fig:pChi}b.

\begin{figure}[tb]
    	\centering
            	\includegraphics[width=\columnwidth]{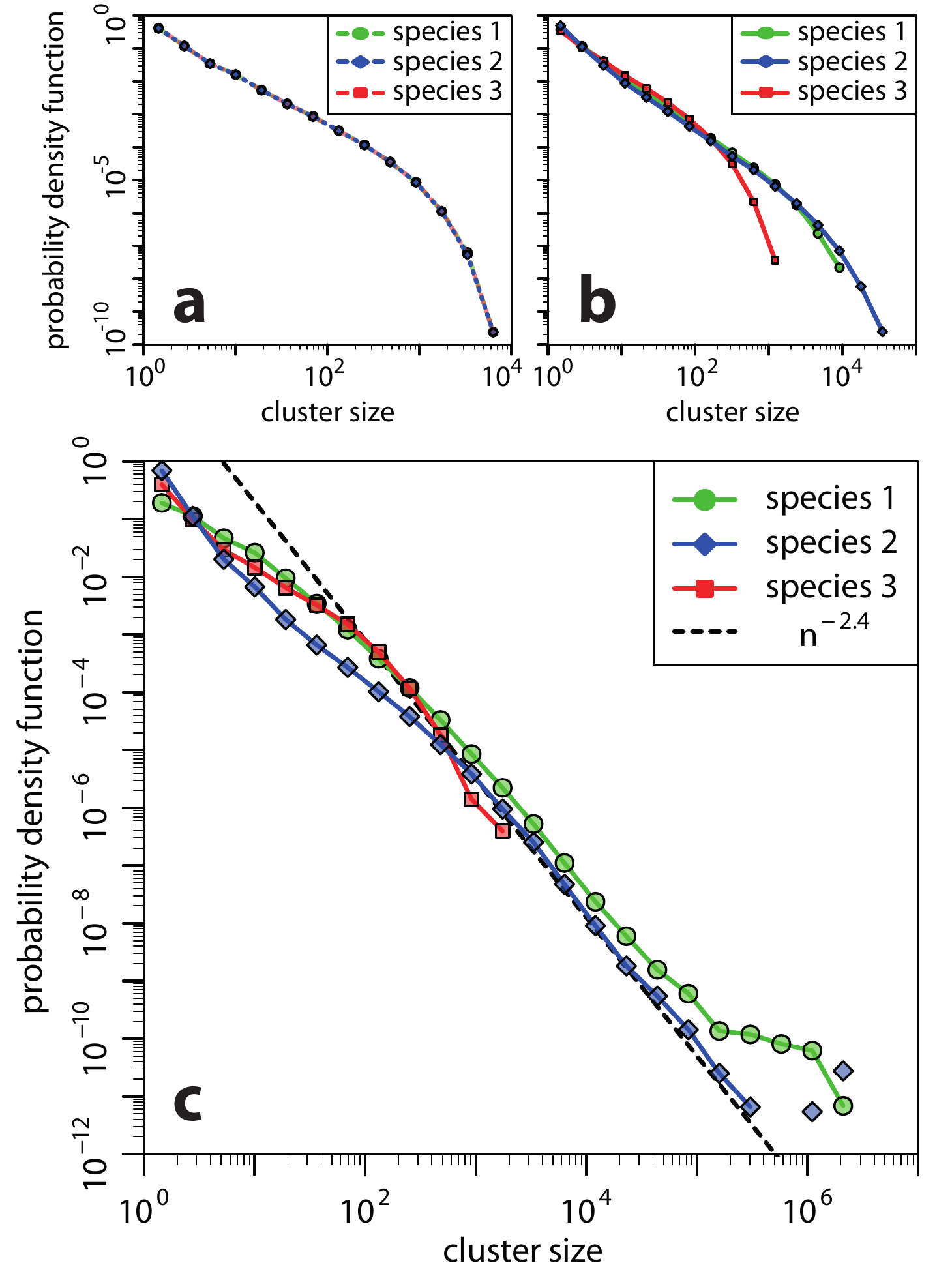}
    	\caption{(Color online) \textbf{Cluster size distributions.} 
\textbf{a} When all species grow at the same rate, all clusters consist of less than 5000 nodes. Here $v_1 = v_2 = v_3 =1$.
\textbf{b} When the growth rate of species 1 is decreased, species 3 becomes less abundant and large clusters of species 1 and 2 become more likely. Here $v_1 \approx 0.5$, and $v_2 = v_3 =1$.
\textbf{c} In the limit $v_1 \to 0$ the cluster size distributions of species 1 and 2 become heavy tailed with a cut-off set by the system size. Here $v_1 \le 0.007$, $v_2 = v_3 =1$. For all plots $L=2048$.
}
    	\label{fig:clusterDist}
\end{figure}

While the value of $\chi$ quantifies the average amount of clustering, it does not provide information on the cluster size distribution. In the case where all species grow with the same rate, Fig. \ref{fig:space}b suggests that clusters have a characteristic size. Indeed, Fig. \ref{fig:clusterDist} shows that the cluster size distribution in this case sharply decreases for clusters larger than $1000$ nodes. When the growth rate of species 1 goes to zero, however, species 3 continues to be organized in small clusters, but large clusters of species 1 and 2 become much more likely. The cluster size distributions of these become exceedingly broad culminating in a heavy tail distribution with a cut-off that is set by the system size.

An alternative approach that has been applied to describe the spatial organisation of the rock-paper-scissors game is the pair approximation \cite{Szabo04, Tainaka}. Here the time evolution of the perimeters $p_{ij}$ are expressed through the probabilities $p^i_{jk}$ of a random node belonging to species $i$ and a random pair of its neighbors belinging to species $j$ and $k$, respectively. When all growth rates are equal, the pair approximation gives the steady state probabilities \cite{Tainaka} 
\begin{eqnarray}
        p_{1} = \frac 13 , \quad  p_{11} = \frac{5}{27},  \quad p_{12} = \frac{2}{27}  \label{pairApprox} .
\end{eqnarray}
Since \eqref{p12tilde} $\tilde{p}_{12} = \frac 19$, the pair approximation predicts $\chi = 1.5$, which is far from the observed value of $\chi = 2.5$. This further illustrates that the incapability of the pair approximation to describe the behavior of the rock-paper-scissors game.

\noindent \textbf{ \textit{  Discussion}}
--- Our results quantitatively describe how spatial clustering slows down the dynamics of the rock-paper-scissors game, and why this leads to a labyrinthine spatial organization in the limit where one species grows slowly compared to the two others. An organization that includes a new type of excitable fronts that propagate on self-organized labyrinthine clusters distributed over many length scales.

In this limit of one slow species, the largest clusters of both the slow species and its prey cover a large fraction of the system, as seen in Fig. 2d.  This consequence of the labyrinthine configuration would not be possible in site percolation, where each of the large species would need to occupy close to $60\%$ of the nodes to percolate \cite{stauffer1992introduction}.

Interestingly, the extreme version of the rock-paper-scissors ecology with one slow species bear resemblance to the forest fire model in a fire-tree-ashes analogy \cite{bak1990forest, drossel1992self, christensen1993self}. The slow species would then be forest, which is burned by fire, which is replaced by ashes, from which trees can again slowly grow. The main differences from existing forest fire models are that in the present system trees can only grow in the neighborhood of other trees and fire can only be extinguished in the neighborhood of ashes. 

The method of quantifying how much clustering slows down the dynamics of a spatial system, compared to the mean field approximation, is quite general, and we expect it to be applicable on a broad range of dynamical systems. In particular, it may be useful predicting the spatial organization in predator-prey models, which continues to attract much attention within the field of complex systems \cite{lugo2008quasicycles, vasseur2009phase}.


\end{document}